\documentclass[conference]{IEEEtran}
\usepackage{graphicx}
\usepackage{textcomp}

\usepackage{subcaption}
\usepackage{dblfloatfix}

\usepackage{flushend}

\usepackage{cite}

\usepackage{amsthm,amsmath}
\usepackage{graphicx}

\usepackage{booktabs}
\usepackage{caption}

\ifCLASSINFOpdf
  \DeclareGraphicsExtensions{.pdf,.jpeg,.png}
\usepackage{url}

\begin{document}

\title{Cybercrime Investigators are Users Too! 
Understanding the Socio-Technical Challenges \\Faced by Law Enforcement}

\author{\IEEEauthorblockN{Mariam Nouh\IEEEauthorrefmark{1},
Jason R.C. Nurse\IEEEauthorrefmark{2},
Helena Webb\IEEEauthorrefmark{1}, and
Michael Goldsmith\IEEEauthorrefmark{1}}
\IEEEauthorblockA{\IEEEauthorrefmark{1}Department of Computer Science, University of Oxford, UK\\ Email: 
\{mariam.nouh, helena.webb, michael.goldsmith\}@cs.ox.ac.uk}
\IEEEauthorblockA{\IEEEauthorrefmark{2}School of Computing, University of Kent, UK \\Email: j.r.c.nurse@kent.ac.uk}}

\IEEEoverridecommandlockouts
\makeatletter\def\@IEEEpubidpullup{6.5\baselineskip}\makeatother
\IEEEpubid{\parbox{\columnwidth}{
    This is a preprint of an article to be published in the Proceedings of the 2019 Workshop on Usable Security (USEC) at the Network and Distributed System Security Symposium (NDSS), 
   24-27 February 2019, San Diego, CA, USA.
   https://dx.doi.org/10.14722/usec.2019.23032 \\
   www.ndss-symposium.org
}
\hspace{\columnsep}\makebox[\columnwidth]{}}

\maketitle

\begin{abstract}

Cybercrime investigators face numerous challenges when policing online crimes. Firstly, the methods and processes they use when dealing with traditional crimes do not necessarily apply in the cyber-world. Additionally, cyber criminals are usually technologically-aware and constantly adapting and developing new tools that allow them to stay ahead of law enforcement investigations. In order to provide adequate support for cybercrime investigators, there needs to be a better understanding of the challenges they face at both technical and socio-technical levels. In this paper, we investigate this problem through an analysis of  current practices and workflows of investigators. We use interviews with experts from government and private sectors who investigate cybercrimes as our main data gathering process. From an analysis of the collected data, we identify several outstanding challenges faced by investigators. These pertain to practical, technical, and social issues such as systems availability, usability, and in computer-supported collaborative work. Importantly, we use our findings to highlight research areas where user-centric workflows and tools are desirable. We also define a set of recommendations that can aid in providing a better foundation for future research in the field and allow more effective combating of cybercrimes.
\end{abstract}



%

\section{Introduction}

With the spread of technology and the global reach of the internet new ways of cyber criminal behaviours are arising every day. According to Mike Hulett, the head of operations at Britain's National Cyber Crime Unit, in 2017 about 50\% of all recorded crimes in the UK involved cyber in some way. Additionally, around 68\% of large UK businesses have been victims to cyber security breaches or attacks~\cite{governmentEuropa}. Law enforcement is facing numerous difficulties trying to keep the pace with the increase in numbers and the evolution of techniques used by cyber criminals. Previous studies suggest that the methods and processes typically used by law enforcement when investigating traditional crimes do not necessarily apply in the cyber world~\cite{brenner2010cybercrime, williams2008catch}. Thus, there is a need to adopt a different process and build a new set of skills and knowledge in order to be able to mitigate these technologically advanced crimes. This is essential mainly as the cyber criminals are usually technologically-aware and are always adapting and developing new tools to allow them to stay ahead of law enforcement investigations~\cite{Wired18, CiscoPolice, nurse2018cybercrime}.

In this article, we study the current practices followed by law enforcement, and more generally security intelligence companies, when investigating cybercrimes and gathering intelligence. We aim to identify the general characteristics related to the investigation process of cybercrimes. This will allow the identification of empirical socio-technical challenges currently faced and areas where new technologies, processes, and workflows are necessary. Identifying these challenges will lead to formulating user-driven requirements for designing and building processes and tools to improve the day-to-day operations of cybercrime investigators. Current cybercrime research, especially those looking at the technical side, tend to primarily focus on proposing and developing new solutions and tools that researchers believe are required by cybercrime investigators. However, little research in the literature focuses on understanding the needs and challenges that are actually being faced by the investigators. The novelty of our work is in attempting to bridge that gap and thereby set a better foundation for future research in the field.

To this end, we conduct a qualitative study using a combination of direct questionnaires and semi-structured interviews. The questionnaires are used to gather generic information about participants role and expertise within the field of cybercrime, while the interviews gather in-depth data. We interview ten experts who work on extracting intelligence and investigating cybercrimes from government and private sectors in the UK. Participants hold varying roles and responsibilities including operational and managerial positions. This gives us diversity in the collected viewpoints  and richness in the data. We use thematic analysis~\cite{braun2006using} to identify common themes and patterns that arise from these interviews, and explore these in the context of our research aim. 

In summary, the paper makes the following contributions to the socio-technical, usable security and cybercrime fields:

\begin{enumerate}

\item It establishes an empirical understanding of some of the key processes used by cybercrime investigators within government (e.g., police and law enforcement) and private (security intelligence) sectors.

\item It identifies a set of outstanding and important challenges (practical, procedural and usability-based) faced by investigators while combating cybercrimes and gathering intelligence. This is useful at directing future research towards more user-centric approaches, practices and systems for investigators.

\item It presents recommendations for areas of improvement associated with the processes of intelligence sharing, reporting cybercrimes, skills and training, and improving the usability of cybercrime systems. 
\end{enumerate}

\section{Background and Related Work}

In this section, we begin with defining what is cybercrime and its main types. We then review related studies on intelligence gathering and policing of cybercrimes.

\subsection{What is Cybercrime?}
There have been several arguments in the literature over the exact definition of cybercrime with no single universal definition~\cite{nouh2016towards}. The European Commission proposed the following definition for cybercrime: ``criminal acts committed using electronic communications networks and information systems or against such networks and systems"~\cite{EuropeanComm}. The definition incorporates crimes that were facilitated by computers and those that were committed against them. Two main categories of cybercrimes exist in the literature: these are computer-enabled, and computer-dependant crimes. The former includes traditional crimes that can be enhanced in scale and reach using computers and networks, while the latter includes crimes that can only be committed using computers and networks~\cite{UKGOV}. For example, a phishing attack and denial of service attack are considered a computer-dependant crime, on the other hand, cyber-fraud and data theft using hacking, key-logging, and social engineering are classified as computer-enabled crimes. 

\subsection{Intelligence Gathering and Cybercrime Investigations}

\begin{figure}[h]
  \centering
  \includegraphics[width=.75\linewidth]{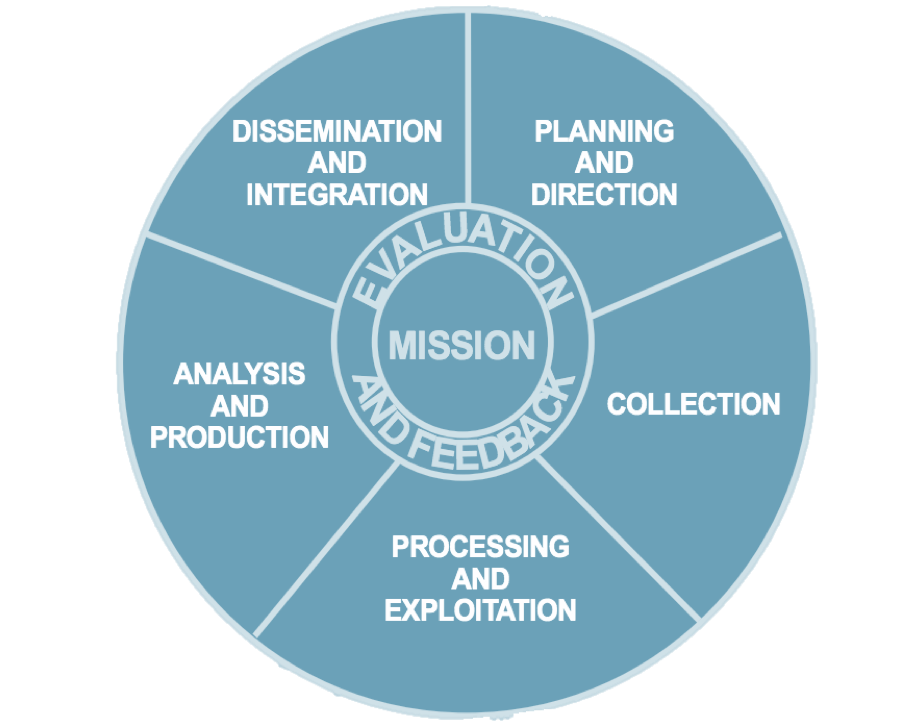}
    \caption{The Intelligence Cycle.~\cite{intCycle}}
    \label{fig:cycle}
\end{figure}

We reviewed existing research related to the process of intelligence gathering. We find that most of the existing literature focuses on identifying different investigation models to guide law enforcement through the investigation process~\cite{Hunton2011,ciardhuain2004}. Although these models may seem different as they use different terminology to define the models' activities, most of them have similar processes. As shown in Figure~\ref{fig:cycle} the intelligence cycle, at its core, consists of six main steps: Planning and Direction, Collection, Processing and Exploitation, Analysis and Production, Dissemination and Integration, and Evaluation and Feedback. The planning and direction step focuses on identifying what are the questions to be answered and plan the best course of action to be followed. The collection step deals with collecting information and data (overtly and covertly) to help answer the identified questions. The processing and exploitation step is when the heterogeneous data is reformatted into a common format for future analysis. The analysis step is the heart of the intelligence cycle, where the processed data is fit together in order to find answers and produce intelligence. The dissemination process is where the analysed data ``intelligence'' is shared with the intended stakeholders. Finally, the evaluation and feedback step is where the stakeholders assess and give their views on the intelligence by either accepting it or coming back with more questions to answer by repeating the process. 

While there are many research efforts concentrated on understanding and analysing the tactics used by cyber criminals \cite{chiesa2008profiling,nurse2018cybercrime,nurse2018bd}, very little research exists that focuses on understanding the processes, challenges, and needs of law enforcement. Majority of the research literature focuses on challenges related to the digital forensics side of cybercrime investigations~\cite{QUICK2014273, quick2017pervasive, LillisBOS16}. Quick et al.,~\cite{QUICK2014273} surveys digital forensic literature on challenges related to data volumes. From the survey, they concluded that there is still a need for further research in multiple  digital forensic areas. In particular, more practical solutions that can be applied within the real-world environment.  

In an effort to understand the effects of technology on policing, the HMIC (Her Majesty's Inspectorate of Constabulary)~\cite{HMIC} published a report to study the current readiness of police services to effectively deal with cybercrimes and their victims. The main outcomes from the study were related to providing training, awareness, and guidance for all those involved with policing cybercrimes. In addition, they recommended raising the level of capabilities within law enforcement in digital forensics and examining digital devices. Another study by Hunton et al.,~\cite{hunton2012managing} focused on identifying different technical investigation roles needed for policing cybercrimes. These roles are technical enquirer, network investigator, forensic technician, digital forensic examiner and technical domain expert. They highlight that when complexity and risk of the cybercrime increase, the level of specialist technical skills and knowledge should also increase. 
Moreover, another study by Harichandran et al.,~\cite{harichandran2016} focused on identifying the needs of cyber forensic investigators. They conducted a survey of participants from different occupations including cyber forensic students, professors, law enforcement, and practitioners. The results from the survey suggest that participants indicated that the main needs include additional funding, advanced tools, better communication, and revised laws.

A number of previous studies looked at police officers perceptions regarding cybercrime~\cite{bossler2012patrol,holt2012predictors,senjo2004analysis}. Bossler et al.,~\cite{bossler2012patrol} investigated how patrol officers' perceived their role in responding to cybercrimes, and their current ability to respond to these offences. The study found that the surveyed patrol officers in the US felt that local law enforcement should not be primarily responsible for handling cybercrime cases. Holt et al.,~\cite{holt2012predictors} looked at identifying predictors of patrol officer interest in cybercrime training and investigation in selected United States police departments. They identify the officers' computer skills as one factor influencing their interest in cybercrime training and investigations. Senjo~\cite{senjo2004analysis} performed an exploratory study to gather police officer perceptions of cybercrimes. The findings pointed that the majority of officers recognized cybercrime as a serious problem. However, the perception of the most common type of cybercrime was different from what has been reported in the literature. Senjo suggests that these perceptions are influenced by mass media depictions and stereotypes. 

We find that the majority of the literature in this area that studies how police officers deal with cybercrimes are based in the US. Very little published literature has investigated elsewhere in the world, including the UK.~\cite{hadlington2018qualitative}. Similarly, the majority attempted to use surveys with close-ended questions as a means to collect data. One advantage of using this approach is collecting answers from a large number of participants. However, a disadvantage is that it limits the ability to capture insights and understand the reasons behind the chosen answers~\cite{holt2012predictors}. Additionally, related literature studied the views of a particular type of participants dealing with cybercrimes (e.g., local police officers). Very few looked at this issue from multiple viewpoints of different participant types (i.e., local officers, regional and national cybercrime units). In this study, we address these gaps by interviewing participants dealing with cybercrimes from government-sector (including local and regional units) and from the private sector. We aim to identify the needs and challenges faced by cybercrime investigators and practitioners using detailed in-depth interviews. In these expert interviews, we do not only focus on a specific aspect of the investigation, such as digital forensics, but we look at the holistic process followed when investigating and therefore touch on several issues related to technical and human aspects.

\section{Methods}

We want to understand the processes, challenges, and needs associated with investigating cybercrimes and gathering intelligence. We look at these issues from the perspective of multiple stakeholders across government and private sectors.

\subsection{Expert Recruitment}
We contacted professionals from different types of organizations, including government and private sectors, since our aim was to gather the perspectives of multiple organizations who deal with investigating and mitigating cybercrimes. Participants were recruited based on their knowledge and experience in investigating cybercrimes and gathering intelligence. We used a snowball sampling approach~\cite{biernacki1981snowball} where we asked initial participants to recommend candidates from their network. The recruited participants held different roles; some possessed technical experience while others held more managerial positions. This diverse sample of experiences and backgrounds was important to capture different perspectives. 

Reaching out to professionals working in this field was a difficult task. Many of the individuals we contacted were very busy and could not afford the time to conduct a full face-to-face interview. In total, we interviewed ten experts six from UK law enforcement, four from the private sector, one of which had ten years of experience working for law enforcement before moving to join the private sector. A similar sample size has been used in previous literature (e.g.,~\cite{obada2017don}). Due to the sensitive nature of the topic, all the participating organisations and individuals requested to remain anonymous.

\subsection{Data Gathering}

In order to gain an understanding of the different socio-technical challenges and needs associated with investigating cybercrime incidents, multiple qualitative methods can be used. Questionnaires, surveys, and interviews are all possible methods that can be used to achieve this~\cite{ritchie2013qualitative}. However, our aim is to understand the problem in-depth and also the needs from multiple stakeholders' perspectives, such as government (law enforcement agents), and private sectors (security intelligence and consultancy companies). Additionally, due to the sensitive nature of this research, it is difficult to gather such detailed information from experts in the field by general means (e.g., surveys) as they are not likely to participate without us gaining their trust and confidence using direct communication. Similarly, relying on observations would give a rich data, however, there are  many difficulties inherent in observing this kind of activity directly. Therefore, we rely on participants reports via a combination of direct questionnaires and interviews to collect our data.

The questionnaire is used to gather generic information about participants role and expertise within the field of cybercrime. Additionally, it acts as a means for us to assess their applicability to participate in the interview process. We then conducted face-to-face semi-structured interviews using open-ended questions. This allowed the interviewees to elaborate on their experiences when dealing with different incidents, as well as providing lengthy and descriptive answers.

The questionnaire focused on multiple themes that aim to give us a general overview of the participants' role, expertise, and technical capabilities. 
After collecting the questionnaire responses, we proceeded with conducting the face-to-face, semi-structured interviews. These interviews ranged from $45$-min to $1.5$ hour long, depending on the availability of the participant and the level of details they were able to provide. We formed a set of predefined questions informed by our review of the literature to guide the interview, but we also asked some probing questions when needed~\cite{turner2010qualitative}. When designing the interview questions we focused on four main themes. These themes are described below.

\subsubsection{Cybercrime Incidents}
The focus of this theme was to understand which types of cybercrime incidents were dealt with by the respective organization, the volume of incidents reported, and whether they had the capacity and resources to investigate all reported incidents. If not all incidents are investigated, we wanted to identify the filtering or scoring mechanisms they used to determine which incidents to investigate. Finally, the method by which victims report cybercrimes and how such interaction occur.

\subsubsection{Investigation Process}
After understanding the types of incidents investigated we then explored the actual process used to investigate a given incident. Under this theme, we collected information on the different tasks performed, if the investigation process was iterative, and whether it was a reactive or proactive process. Understanding the details of the investigation process will allow us to map the different steps performed and, identify where applicable, how each step can benefit from technology adoption. 

\subsubsection{Investigation Team}
This theme sought to understand the dynamics within the investigation team. How big is the team that works on any single case? What role does each team member typically have? Are they all located in the same location? 
Such questions will allow us to ascertain if there is a need for processes, systems or tools to support collaborative investigation sessions. This relates to the field of Computer-Supported Collaborative Work (CSCW).

\subsubsection{Tool Support}
This theme covered the different tools currently being used by  investigation teams. The aim was to understand what were the advantages gained from using these tools, the perceived usability of the tools, their availability, and effectiveness at supporting the investigation tasks. We also ask about the perceived limitations of the tools, which is important to be able to enhance these limitations and give better support to investigators.

All interviews were audio recorded and then manually transcribed producing transcripts for each participant discussion. An ethical approval for this study was granted by the central university research ethics committee. Also, we ensured ethical handling of collected data through an informed consent process for participants, and anonymisation of published results. Additionally, to ensure the validity of the questionnaire and the interview questions, we discussed them with a subject matter expert and incorporated appropriate feedback before conducting the study.

\subsection{Data Analysis}
The data collected from the questionnaires consist of open-ended text (e.g., participant's role) and categorical data (e.g., types of cybercrimes). The categorical data was analysed using descriptive statistics by calculating the frequencies for each category. The results from the questionnaires played a role in guiding the discussion during the interview session to areas that match the participant's experience. 

To analyse the interviews data, thematic analysis was chosen as it best fits the exploratory nature of this study. It allows for theoretical freedom, as it provides a flexible and useful research tool, which can provide a rich and in depth account of the data~\cite{braun2006using}. Thematic analysis focuses on identifying common themes within the data. A theme is defined as a recurrent feature or topic describing particular perceptions and experiences relevant to the research questions~\cite{QualAnalBook}. We adopt thematic analyses with a mixture of deductive and inductive approach to identify themes and analyse the interviews data~\cite{QualAnalBook}. While deductive approach allows us to guide the flow of analysis according to our aims and predefined generic themes from the literature, the inductive analysis allows us to incorporate additional themes that are new and arise from the data~\cite{fereday2006demonstrating}. This allows us to generate findings from the bottom up in order to be inductive and identify patterns based on participants’ expressed reasoning. As opposed to using rigid pre-set criteria, which may or may not map on to the participants’ perspective. 

We began our analysis with initial coding using a deductive approach by coding the data to a pre-defined set of codes relevant to our research interest. These codes correspond to the four themes described earlier, i.e., cybercrime incidents, investigation process, investigation team, and technology. Then through an iterative analysis approach the coder immerse themselves in the data and assign new codes where appropriate to the recurring ideas. Using this approach makes the findings more robust as it allows us to test emerging themes against new data. The coding process was carried out by a single coder, and the analysis was performed using a qualitative analysis software package called NVivo\footnote{https://www.qsrinternational.com/nvivo/home}. 

An example of our coding process is shown in the quote below, where the participant was asked about their impression of utilising automated tools in their investigation of different cybercrimes. This question falls under the technology and tool support generic theme.  

\textit{``I think fully automated tools need to be introduced into our work. Purely because you know budget cuts. So there's going to be no more resources. So we need these tools, scripts, data mining tools to go through everything we see.''}

We initially assigned the code \textit{impression of automated tools} to highlight the participant's views on the topic during the first coding round. Then in our subsequent iterations,  we assigned the code \textit{drivers for automation} and \textit{resources availability} as they also mentioned the lack of resources and budget cuts as reasons for their views. Overall, we identified $35$ codes in our data, that were then refined and grouped into themes.

\section{Results}
Through the analysis of the interview data, we attempt to understand different aspects connected to the process followed when investigating different cybercrimes. Several themes appeared across the interviews data that show how private and public sectors carry out their investigations. 
In this section, we first present the demographics of interview participants. Second, we discuss results from the collected questionnaires. These give us an idea regarding the  expertise of the sampled participants. Third, we report results from the interview analysis relating to the different identified themes.

\subsection{Participants Demographics}
We interviewed $10$ participants who are experts in cybercrime investigations. The participants worked in government ($6$ participants) and private sector ($4$ participants). All participants were based in the United Kingdom. Of the participants, $4$ were intelligence analysts; $2$ were intelligence researchers; $1$ was a sergeant; $1$ was a detective; $2$ were law enforcement agents dealing with cybercrimes ($1$ is currently working as a consultant). The participants belonged to $5$ different organizations, $2$ private sector, and $3$ government sector. The government sector organizations tackled cybercrime at different levels, including local, and regional forces. This allowed us to collect the viewpoints from experts working at tackling cybercrimes from different levels. The private sector organizations deal with security investigations and intelligence gathering from open and closed sources to provide clients with threat intelligence. 

\begin{table}[ht]
\caption{Participant Groups}
\label{Participants}
\resizebox{\columnwidth}{!}{%
\begin{tabular}{@{}lll@{}}
\toprule
\textbf{\begin{tabular}[c]{@{}l@{}}Group ID\end{tabular}} & \textbf{Roles} & \textbf{\begin{tabular}[c]{@{}l@{}}Organization\\ Type\end{tabular}} \\ \midrule
G1  & Intelligence analysts and researchers & Private and Ex-Gov  \\ \midrule
G2  & Cybercrime officers and detectives & Government \\ \midrule
G3 &  Senior intelligence analysts and researchers & Government  \\ \midrule
G4  &  Senior cybercrime officers and detectives & Private and Government  \\ \bottomrule
\end{tabular}%
}
\end{table}

To ensure proper anonymization and to allow for better protection of our participants identities we grouped them into $4$ groups based on their roles and level of experience. Table~\ref{Participants} summarizes the different groups by providing the Group ID (GID), the job roles of participants, and the type of organization they have worked in (i.e., private, or government).
   
\subsection{Questionnaires Results}
\begin{figure}[t]
    \centering
    \begin{subfigure}{\linewidth}
    \centering
        \includegraphics[width=\columnwidth]{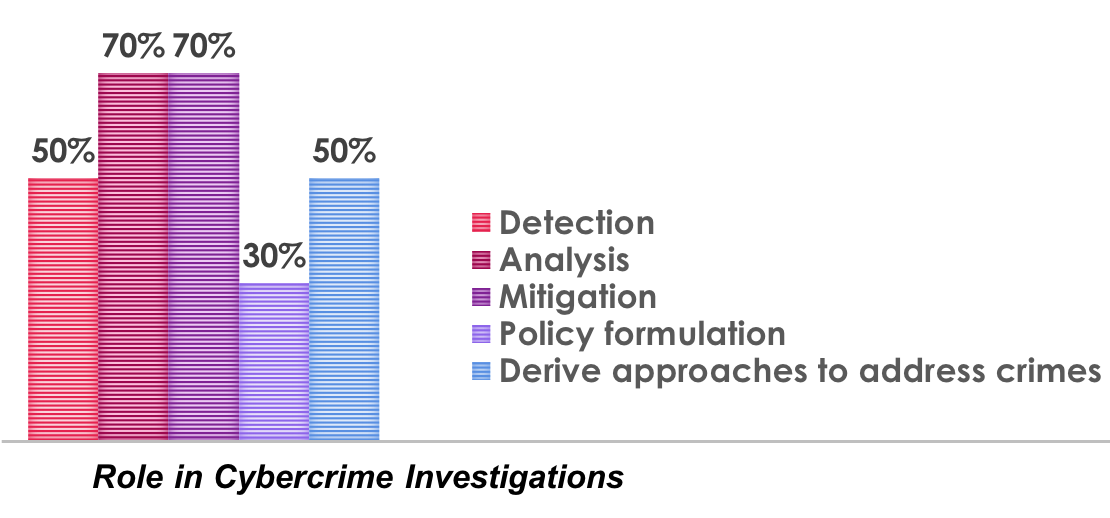}
    \end{subfigure}\\
    \begin{subfigure}{\linewidth}
    \centering
        \includegraphics[width=\columnwidth]{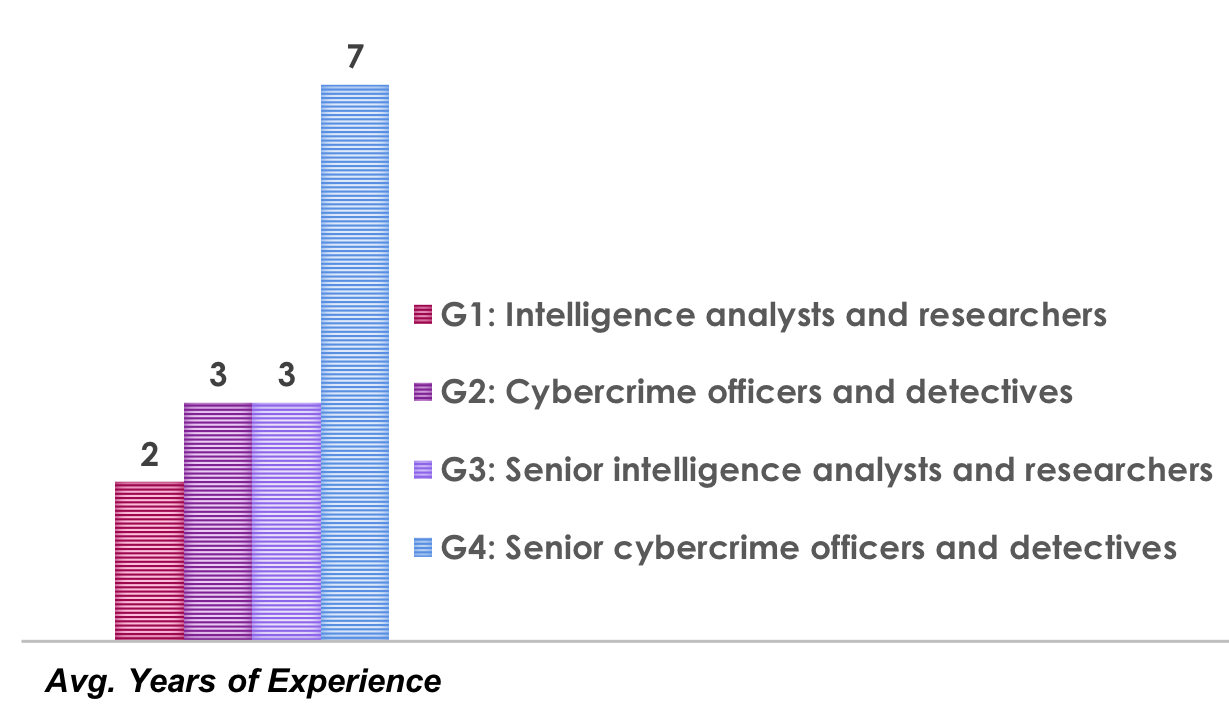}
    \end{subfigure}\\ 
   
    \begin{subfigure}{\linewidth}
        \centering
        \vspace{20 pt}
        \includegraphics[width=\columnwidth]{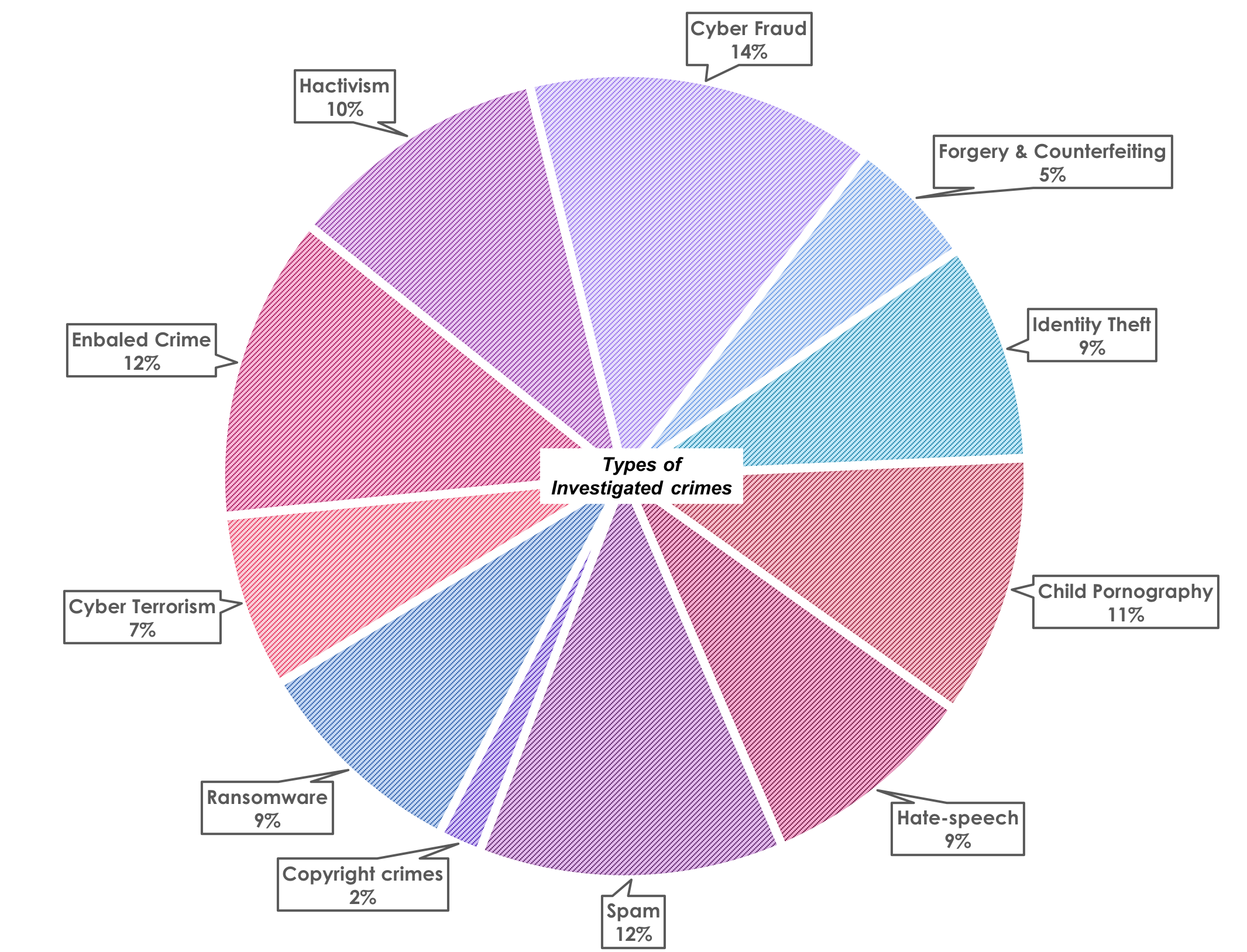}
    \end{subfigure}\\
    \caption{Participant Experiences with Cybercrimes}
    \label{fig:experiences}
\end{figure}

The questionnaires were designed to collect data that can paint a picture of the participants' experience, role, types of cybercrimes they investigated, and their technical capabilities. We summarise the findings below and present a digest in Fig~\ref{fig:experiences}.

\begin{itemize}
\item \textbf{Years of Experience:} The average years of experience participants had was $3.5$ years working in investigating cybercrimes. The minimum was $1$ year and the maximum was up to $10$ years of experience. They held roles that ranged from intelligence analysts, consultants, lead analysts, researchers, and investigation officers. This generates multiple viewpoints and allows us to gather requirements from different angles.

\item \textbf{Types of Cybercrimes:} A list of different types of cybercrimes were presented to the participants and we asked them to select the crimes that they have experience dealing with. The results were distributed between cyber-enabled and cyber-dependent crimes, with the majority having experience dealing with cyber-fraud, spam, hacktivism, and child pornography. Additionally, participants noted that after the 2017 WannaCry ransomware, the numbers of reported ransomware spiked and became the most reported cybercrime.

\item \textbf{Role of Investigation:} We asked if their involvement with the investigation was part of the (1) detection phase, (2) analysis, (3) mitigation techniques, (4) policy formulation, or (5) derive approaches to address crimes. The majority (around $70\%$) were mainly involved in the analysis and mitigation of cybercrimes. Additionally, around $50\%$ of participants were involved in the detection phase and in deriving different approaches to address these crimes. Only $30\%$ of participants had a role in formulating cybercrimes policies.

\item \textbf{Techniques Used:} We asked the participants about the techniques and experiences they adopt when investigating cybercrimes. Among the most common techniques were OSINT (Open Source Intelligence), social network analysis to study relationships and interactions between criminals, and visual analytics. None of the participants reported that they use machine learning approaches in their investigations, and only some reported using data mining techniques.
\end{itemize}

\subsection{Understanding the Investigation Process}
When asked about the process of intelligence gathering, the majority of the participants agreed with the  six steps of the intelligence gathering cycle described in the literature~\cite{ciardhuain2004}. 
However, the initiation of the intelligence gathering process differs between the private sector and government sector participants. This is due to the difference in nature between the two sectors. In the case of the private sector, typically they work with a client and start by collecting and understanding the client's requirements. Thus, they are able to scope down the task of gathering specific intelligence to answer clear requirements. The process tends to take a top-down approach where they formulate a hypothesis and then collect data to prove it. 

On the other hand, with law enforcement, the investigation and intelligence gathering are initiated when they receive a report that someone has been a victim of a cybercrime. In such a case, the intelligence gathering process begins with a bottom-up approach. Meaning they start from the collected data, analyse it, and then try to formulate a hypothesis of what happened based on the analysed data. Although the majority of participants from the government sector emphasised this ``bottom-up'' approach, one participant stated that they tend to use mixed approaches. They may start with a hypothesis or initial belief ``bias'' based on their experience and then look at the data to either prove or dismiss the hypothesis.

\textit{``What I've done over time is, you will have your own kind of biases or stereotypes in your head which guide you. Not in like negative stereotypes but ones that help you. You've just learnt from how you've gone before.''[G3]}

When we asked participants about the differences between investigating traditional crimes and cybercrimes, we received a diverse set of opinions. Some felt that for cyber-enabled crimes there are no differences and the same kind of detective tactics would work. To them, it is still just a crime that has been committed regardless of the means used. Thus, the investigation process is basically the same.

\textit{``The fundamental thing is the majority of crime associated with fraud or cyber is enabled crime.  In other words, it's traditional crime that is just leveraged through a digital device. Nothings changed, it's the existing crime that has been committed for centuries if not thousands of years.''[G2]}

On the other hand, others felt that there are key differences that the cyber aspect brings to the investigation. The main differences mentioned by participants relate to the globalization of cybercrimes, which restricts  how far the investigation may go due to jurisdictional issues. Another key difference mentioned by participants is that cybercrimes compared to traditional crimes tend to have less investigation leads 90\% of the time. This may be due to the sophistication of cyber criminals in covering their online traces, or due to the lack of cyber awareness from the victims. For example, they may accidentally  destroy digital traces and evidence, or report the wrong information to the police.

When investigating a cybercrime incident, participants reported that there are several factors that affect how this process is carried out. These factors include, the type of crime committed (cyber-enabled, or cyber-dependant), who the victim is (individual, small/medium business, large enterprises), type of potential suspects (individual hacker or organized group), what information or intelligence is available (investigation leads), and existing resources available (human and technological resources).

We asked participants about the use of OSINT during their investigations, and the general consensus across participants from different organizations was that OSINT is an important tool for intelligence gathering.

\textit{``Open Source Intelligence is a massive tool for policing, and it's used across the board, not just by us, it's used for all sorts of things, so it is a technique that detectives are familiar with.''[G4]}

Participants stated that its importance comes from the fact that suspects spend a lot of time online and have a large presence either in social media or on the internet in general. One participant stated that OSINT is not only valuable to the cybercrimes investigations but even traditional crimes would have benefited from using online sources to help with the investigation.

\textit{``With some of the historic crimes we investigate, I think we should have been doing more open source research side.''..
``I think open source intelligence has got a much bigger role to play than any more traditional crimes and can be much more successful, we can get a lot better results''[G2]}

\subsection{Collaboration and Investigation Teams}

Participants across different organizations described similar team structure and dynamics when investigating cybercrimes and gathering intelligence. Typically, a given case will have a lead investigator and may have the assistance of an analyst and a researcher to support the investigation. Researchers are responsible for conducting regular monitoring of different sources and preparing intelligence briefs for analysts/investigators or clients. Analysts are responsible for developing the intelligence products, identify trends related to specific expertise and themes in cybercrime. For instance, taking the volume data and make it presentable and useful either to the lead investigator (and the court later on) or to the client. Depending on the size of the department and the available resources, the team may also include technical staff that examine devices whether that is for victim or suspect. Once these are reviewed, all the case data go to the analyst who tries to evaluate it, challenge it, and pick it apart.

We asked participants about the typical technical and cyber experiences of people working in each role. Most mentioned that investigators and analyst do not necessarily come with previous technical knowledge or background, however, they would learn on the job the technical skills needed. This phenomenon was true across all types of organizations, be it private sector or government and police force. 
  
\textit{``Some before they become cybercrime investigators, they were all probably 6 to 12-year detectives, so they've been dealing with burglaries and things like that. But there were a couple that had programming backgrounds, and one had worked in information security.''... `` The basis of it was getting good detectives and good research and analyst skills and then try to give them enough technical knowledge to do what they need to do.''[G1]}

Moreover, we wanted to understand how the collaboration across departments and/or organizations happen when investigating a specific case. In the context of the private sector, this was easily done through emails and online or in-person meetings with people from different departments. Depending on the size of the organization, teams would either all be based in the same office, city, or they may be working with teams based overseas. Either way, they reported that information sharing within and across teams was quite seamless and did not present a struggle to their investigation process. Overall therefore, these represented usable and efficient workflows.

On the other hand, participants from the government sector had a different and more challenging user experience. The flow of information and intelligence sharing from different cybercrime units and agencies was a challenge. Participants from different cybercrime force units mentioned that they often have difficulties connecting information from a case they are investigating with possible linked cases in other forces. 

\textit{``So very little intelligence sharing happens within cybercrime. ... Getting stuff up from forces to regional units and vice versa is a lot more difficult. ... Say you had an offence down south [placeName] and you had an offence up north [placeName] and they were connected, whether it be like a suspect email address, the chances of connecting those two offences, depending on what level it came in, could never happen basically.''[G3]}

This is particularly true for information sharing between units at different levels (local force and regional force) as well as between force units at the same local level. However, the information sharing is currently better between force units within the regional level. Moreover, one participant mentioned that sometimes such linkage of cases may happen by chance.

\textit{``Because you have such a workload that you'd like to be most of the time now identifying that something connected to something else is purely based of an individual's memory as opposed to a system alerting. So those are the kind of next steps we need.''[G3]}

Another participant expressed the need for a connected system across forces that is able to calculate such associations between relevant cases. 

\textit{``If we had the ability to share our data from our jobs and their jobs and we had the analytical ability to be able to pick out and go hang on a minute, their, our job links to that one in [placeName]. Sometimes that linkage happens by virtue of us having conversations but not you know automated.''[G4]}

\subsection{Technological Capabilities and Tool Support}
Within the general theme of technological tools and capabilities, participants reported that multiple tools are usually used during investigations. Each of these tools is used to achieve a specific task, such as data collection, analysis, or visualization. When asked about the interoperability and interactions between these tools, the majority stated that these tools do not work well together. Instead, it requires significant  manual effort to extract the output of one tool and map it into another tool for further analysis. 

\textit{``They're not necessarily as good at each part of the investigation. Certainly, most of them have a focus at one point of the investigation cycle. And then the rest might be add-ons which are getting better but do not necessarily, see the job all the way through as good as you'd like.''[G1]}

Moreover, participants stated that sometimes they work on different systems and networks, due to some corporate regulations some tools are only allowed to reside within public systems. Thus, different tools will be forced to sit on different networks. 

\textit{``They very much work in isolation, in most cases you would use a tool, and say it was going to be the main tool that I was using to build court products, it is great for visualisation but there is very little in terms of labelling, exporting, producing reports. So I might do all my working outside that tool, narrow it down to the picture I want to show and then manually map that in the tool.''[G3]}

Participants from the government sector also mentioned that there is currently no national mandate for the selection of tools to be used within forces. Therefore, each police force operates a different system and process, which adds to the challenge of sharing knowledge and intelligence between forces. This suggests a much larger issue with regards to creating an efficient, effective and ultimately widely usable cybercrime investigation process.

\textit{``Tools are not mandated nationally to each force, and if it was a national infrastructure for a lot of these things then that would be really helpful. It's generally on a force by force basis.''[G4]}

We asked participants from government sector about how the current police systems are being used to report cybercrimes. Participants expressed that current police reporting system for cybercrimes (called Action Fraud (AF)~\cite{AF}) suffers from a number of issues. AF is the UK's national fraud and cybercrime reporting centre, which was initially designed for reporting fraud crimes specifically. The system was not designed for recording cybercrimes and thus generates ambiguity for the AF system users ( police officers or self-reporting victims) when reporting these crimes. This leads sometimes to offences being recorded under the wrong category. 

\textit{``It was not really built to deal with cybercrime, and so they had real issues in kind of categorising offences, and when they get reported you have the victim knowledge being able to accurately report it, and then we have the call taker knowledge.''[G4]}

Moreover, many details specific to cybercrime data are not properly recorded and stored. This is due to the lack of structured methods to collect the necessary cyber-related information, such as IP addresses, bitcoin wallets, usernames, and so on, which makes reports collected from different victims differ in the level of detail. This has an effect on how these reports are being searched and eventually leads to missing links between crimes that may share similar details, such as email or IP address. 

\textit{``The police reporting systems themselves are not even necessarily catered to record cybercrime offences.  They have to be kind of pushed in under a different category sometimes and systems might be set up to record things like home addresses, vehicle indexes and everything like that, but very little systems are set up to record email addresses, IPs and everything like that.''[G3]}

Additionally, participants mentioned that there may be a shortage in the number of call-takers within Action Fraud, which has an effect on the number of cybercrimes that gets assigned to different forces. Participants also mentioned complications related to the limited level of cyber knowledge and experience in dealing with cybercrimes that initial call-takers have. This leads them to misdirect crimes to the wrong department which causes delays in responding to the crime. Also, one participants perceived a link between how prevalent cybercrimes are in the news to the number of crimes that get communicated to them from the call-takers.

\textit{``One of the issues that we've identified within our force is when a victim calls the Police, they get a really patchy response as to how it's dealt with because our call takers have a lot of experience of dealing with reports of rapes and assaults but when it comes to cybercrimes they do not really know what to do so it often gets misdirected or delayed.''[G2]}

We talked with participants about the implications of using technology to automate certain investigation processes. Many of them had some concerns regarding the automation effects on the process, and expressed preference to perform investigation tasks manually. Either due to lack of technical skills, to re-validate previously obtained results, or for the fear of missing some evidence that is relevant to the context they are investigating. Nevertheless, participants from both private and government sector emphasised that automating certain processes may be necessary, but it should be done under human control and guidance.

\textit{`` Automated tools can be useful for finding new data. I still think, and may be it's a cultural thing, but it still requires very human element to it. I know the context of what we're looking for and I can interpret the messages and the chats that I'm reading and pick up different things and bring it all together. I might see something in one investigation, actually I realize is relevant for another one.''[G1]}

Including the human (user) in the process is particularly important in order for them to understand the results produced by the tool, and be able to trace back the process from the input to the produced output. This is important because investigators need to explain their findings to clients and to be able to defend it in court.

\textit{``You do not want to use a tool and just feel like you are just clicking a button'' ... ``If you are stood in court you want to be able to explain what is happening behind that tool.''[G3]}

We asked participants about what technical capabilities and desirable tool features they think will benefit the way they carry their investigations and gather intelligence. Government sector participants all agreed that they need a substantial upgrade in their IT infrastructure to be able to cope with the sophisticated cybercrimes they investigate. This was not an issue for the private sector. Additionally, both groups described the need for more user-centred tools to allow them to easily do their jobs, tools with better support for data visualizations, and tools that can aid in analysing bulk amounts of online data to cope with the evolving online ecosystem.

\textit{``We could be so much better with better technology, share data, analyse data, be proactive with data. So an improvement in technology would certainly assist us for the future. I think we do things perhaps the long way and over time we could improve on that.''[G4]}

In summary, this section highlighted the main findings gathered from the experts' interviews. These are mainly spread across topics related to the process of cybercrime policing, the collaboration between investigation teams, and technological capabilities of the investigators.

\section{Key Challenges Faced by Cybercrime Investigators}

The literature from the criminology field discusses a number of challenges facing the police when investigating crimes. For example, some of the discussed challenges are related to police patrol and organising hot-spot policing in order to minimize numbers of street crimes~\cite{hutt2018data}. Moreover, Ratcliffe~\cite{ratcliffe2008knowledge} argues that there is a need for a new shift in policing from relying on old knowledge, which is relating to the criminal activity that is typically collected in traditional crimes, to a new knowledge, which is focused on the crime event that is collected from public open source information. 

While some of the challenges that are inherent within the traditional crimes field transcend to cybercrimes, a number of new challenges emerge that is linked to the cyber-world. For example, cybercrimes are borderless by nature, thus there are several legal and jurisdiction issues related to investigating cybercrimes. In traditional crimes, the offender has a physical link to the crime scene which makes the investigation and detention easier. However, in cybercrimes, this is not the case. Offenders may be in a different place, state, or country from the victim. This makes cooperation across different agencies a necessity. 

Through the analysis conducted in this study we are able to identify some key socio-technical challenges that cybercrime investigators face. These challenges emerged from the themes discussed previously and are grouped into four main topics: (1) Reporting of Cybercrimes, (2) Information Sharing, (3) Tools and IT infrastructure, and (4) Skills and Technical Abilities. We discuss these further below.

\subsubsection{Reporting of Cybercrimes}

One of the main issues is related to AF system and how it was initially designed and used. The system users faces many ambiguities when recording the reported crimes, which results in miss-categorizing cybercrime offences. This leads to inconsistencies in the level of details collected from victims, which typically depends on how technologically-aware the victim is and the level of cyber-experience held by the call-taker preparing the report. There is also an increased likelihood of misdirecting the report to the wrong department thus causing delays in addressing the crime. It also increases the time and effort needed to conduct the investigation. Additionally, the lack of structured methods to collect the necessary cyber-related information, has multiple implications on the investigation process, such as missing some connections between possibly related victim reports. Overall, a major review of the questions included in AF was felt to be needed and appropriate training is necessary to better equip call-takers to handle cybercrime related reports.

\subsubsection{Information Sharing}
One of the most significant challenges faced by law enforcement is the lack of centralised coordination of intelligence sharing between forces and agencies working on cybercrimes. This has multiple effects, firstly, the intelligence products produced by forces may be incorrect since they do not have the big picture and are not fully aware of what crimes are being reported in different regions. Second, the lack of communication between different local forces results in missing possible connections between different reported cybercrimes. This makes establishing links between cases a real challenge. Third, following up on specific cybercrime cases and providing status updates either to the victim or the media when requested, is difficult due to the lack of sharing case information. This has a negative effect on the police reputation and image as it may make them look incompetent in the eyes of the public. Moreover, given that cybercrimes have a border-less nature, unlike traditional crimes, they may involve suspects in distributed locations. Therefore, having better communication, coordination, and data sharing between different units and forces will have a positive effect on mitigating and responding to these crimes.

\subsubsection{Tools and IT Infrastructure}

A key challenge associated with investigating cybercrimes is related to the level of IT infrastructure available to police forces and the cyber capabilities of investigators. In recent years there have been several budget cuts that prevent forces from investing in upgrading the IT infrastructure and acquire advanced tools. This limits their capabilities and has effects on the efficiency and quality of the conducted investigations. Similarly, investigators expressed that a lot of the investigation time is spent doing manual tasks that may be saved by utilising some automated or semi-automated tools. Examples for such tasks include looking at OSINT data to collect intelligence,  analysing terabytes of data from server logs and victim devices. Furthermore, the lack of a national mandate of tools and systems to be used in cybercrime investigations adds another challenge for collaboration between different cybercrime units. Police forces are generally decentralized in the UK. While having a structure of separate 43 police forces~\cite{independent2013policing} may be fit for policing traditional crimes, this may not be the best structure for delivering effective actions against crimes that are cross-jurisdictions (i.e., cybercrimes). This structure led each cybercrime unit to become a silo, operating a different set of processes and tools and unable to interoperate. Similarly, this interoperation issue also exist for the different tools used, where a lot of the investigation time is spent in manually modifying and formatting data to be able to move it between different analytical tools.

\subsubsection{Skills and Technical Abilities}

Another challenge is related to the cyber skills of the staff. There is a lack in the number of skilled technical personnel working on investigating cybercrimes. The majority are highly experienced investigators who have been working with traditional crimes and are now, because of the increasing numbers of cybercrimes, are moving toward the cyber field. Investigators are highly trained in gathering intelligence and investigation tactics but might not be as well-trained in the cyber-world. There are, therefore, open questions pertaining to employee training and expertise. This is also unlikely to be a problem only faced in this geographic area due to the widespread increase in cybercrime. Potential avenues that may be explored going forward include further training and upskilling for the necessary personnel, but also creating tools that are better geared to supporting user skills and activities. The ideal case will be the provision of practices and tools that are easy to use and would reduce the learning time for new cybercrime investigators.

\section{Conclusions and Recommendations}

In this study, we interviewed participants from both private and government sectors on their experience with investigating different cybercrime incidents and the process of gathering intelligence. The interviews were focused on four main themes: (1) investigation process, (2) cybercrime incidents, (3) collaboration and information sharing, and (4) technology. Overall, both groups agreed on the steps needed to gather intelligence; starting with identifying the problem to address, collecting relevant data and information,  processing and analysing the collected data to identify links and patterns, and finally dissemination step where the created intelligence is shared with appropriate stakeholders. These steps are similar to what is discussed in the literature related to the investigation and intelligence gathering cycle~\cite{Hunton2011,ciardhuain2004}. Although both groups agree on the general process, there exist some differences in the types of data they collect and use in the investigation. For example, when gathering OSINT, law enforcement is more restricted with the type of data they can collect and use in their investigation. Participants mentioned that when tracking people communication in online forums, any forum that requires a login credential to access would be off-limits. Even if they did access it, this data could not be used as part of the case and will not be used in court. 

With regards to the second theme, collaboration and information sharing, although this happens seamlessly between different departments within a single private sector organization, sharing across organizations is not required as they do have competing interests and do not tackle any national issues. On the other hand, for government and law enforcement sector it is completely the opposite as little information sharing happens within local forces and across forces. Finally, for the technology capabilities, again we see diverse responses between the two organization groups. The private sector seems to be more technologically advanced by having the supporting IT infrastructure, the advanced tools, and the skilled human capital. This might be expected due to issues around remuneration and specialisms, but has significant implications for cybercrime investigations within law enforcement.

\subsection{Recommendations}

There is a number of challenges currently faced by cybercrime investigators working on government and private sectors. Some of these challenges are inherent in the area of policing traditional crimes and have transcended to cybercrimes. Others are new challenges that emerged because of the way the cyber space is designed. To overcome these challenges, we provide the following recommendations:

\begin{itemize}

\item Cybercrimes reporting systems: Cybercrime victims can use the Action Fraud (AF) online system to create a report online, or they can report the crime by phone where the report is created by a call-taker. Addressing the issue of the poor quality of reported information in the government sector that leads to misdirection of some reports to the wrong department, a review and an update to the current questions included in AF online system is necessary. Moreover, a study and evaluation of the usability and user experience of the online AF tool is needed to identify the sources of ambiguity and areas for improvement. Additionally, it is necessary to provide a way for updating and tracking the status of each reported crime during the investigation cycle, while providing AF call-takers and other relevant users access to such updates. This will allow them to provide prompt updates to any enquiring victims.

\item Information sharing: A centralised coordination of intelligence and partnership is required at a national and/or international level to be able to share intelligence and make effective use of resources. Having a central system connected across forces will allow a seamless flow of information between local, regional, and national agencies dealing with cybercrimes. To achieve this, further research is needed to facilitate the exchange of intelligence in a secure and usable manner across different systems and cybercrime units. Additionally, it is critical to study and evaluate the best architecture for information sharing across different levels of cybercrime units; common options include a hierarchical architecture (i.e., local unit connected to regional, and regional to national) or a flat architecture (i.e., having each cybercrime force unit connected to the others). Additionally, designing appropriate access control models is needed to facilitate the availability of intelligence to those who require access to it while maintaining the confidentiality and integrity of that information.

\item Tools and IT infrastructure: Current tools used within the investigation process as suggested by participants from both government and private sectors, sometimes lack an important requirement that is related to the transparency of the analytical process. Investigation and intelligence gathering tools should be designed with \textit{human-in-the-loop} concept in mind. As an investigator, he/she needs to be able to understand how the tool is working, and how particular results were reached. This is critical to them as they need to be able to explain their findings and defend them in court and in other legal settings. To achieve this, the HCI and usable security research community need to investigate how  best to improve the user interaction and involvement with intelligence tools. Additionally, the systems and tools used in each cybercrime force unit should be mandated nationally. This will facilitate a common infrastructure of sharing of analytical capabilities and linked intelligence. Furthermore, this will increase the interoperability between different tools and systems, and reduce the amount of incompatible tools currently being used across different organizations. Finally, an investment in upgrading the current IT infrastructure in different cybercrime force units is necessary, to allow investigators to have the capacity to analyse the vast amounts of data being generated from these crimes.

\item System usability and human experiences: Many of the investigators working in cybercrime from private and government sectors come with experience in investigation tactics associated with traditional crimes, and may lack the technical skills. One way to address this, is to provide more training, capacity building, and awareness in cyber for law enforcement forces (especially at the local-force level). Possibly a more important point however is in improving nature of systems and workflows (individual or collaborative) to make them more user-centred and built to suit individuals' tasks. Increasing the usability of the cybercrime investigation and intelligence systems is crucial as it can reduce undue burden on the technical abilities of workers. The ideal situation is for tools and workflows to be designed to take advantage of investigators' expertise and support them in cases where this may need bolstering.

\end{itemize}

We believe that future research within this area needs to focus on exploring these topics. In our future work, each of these recommendations will be further explored and detailed with the aim of providing a suitable framework to implement them. This study is an exploratory study that provides a starting point that exposes some assumptions and challenges faced by practitioners. Additionally, given the small participant sample, more work needs to be done for generalizability of the findings. One way to achieve this is to utilise these findings to aid the formation of questions that can then be put into a survey and sent out to a wider audience. 

\subsection{Study Limitations}

There are a few limitations to our research which should be noted.
When conducting semi-structured interviews there are unavoidable variations in the level of details gathered from each participant. Additionally, in order to understand the problem in-depth and capture the views of participants with their own words, we chose to conduct face-to-face interviews. However, this limited us in the number of participants that we were able to reach especially given that the targeted community is considered hard-to-reach for academics. Similarly, the sampling of participants was chosen from different agencies (government and private) who deal with cybercrime investigations. From the government agencies we were able to gather the perspective of local cybercrime task forces and regional cybercrime units. However, we could not get access to participants at a national cybercrime agency level. In future work, it would be interesting to include the views and perspectives of experts from national cybercrime agencies, as well as expanding our sample size to allow even more insight into the problems faced.

\section*{Acknowledgment}
The authors would like to thank the Ministry of Education and King Abdulaziz City for Science and Technology (KACST), Saudi Arabia for financially sponsoring and supporting Mariam Nouh's PhD programme. We also thank all participants for their valuable time and for participating in this study. Finally, we thank the anonymous reviewers for their helpful comments.

\bibliographystyle{unsrt}
\bibliography{ref}

\end{document}